\documentclass[ amsmath,amssymb,nofootinbib, prd]{revtex4}
\usepackage{graphicx}
\usepackage{amsmath, amssymb, graphics, epsfig}
\usepackage{dcolumn}
\usepackage{bm}
\usepackage{hyperref}
\usepackage{epstopdf}

\newcommand{\nc}{\newcommand}
\nc{\ba}{\begin{eqnarray}}
\nc{\ea}{\end{eqnarray}}
\newcommand\be{\begin{equation}}

\newcommand{\calH}{{\cal H}}
\newcommand{\calR}{{\cal R}}
\newcommand{\calP}{{\cal P}}
\newcommand\mPl{{M_{P}}}

\newcommand{\bfk}{\mathbf{k}}

\newcommand{\cs}{{c_{s}}}
\newcommand{\bfx}{\mathbf{x}}

\newcommand{\bmk}{\mathbf{k}}

\preprint{}

\begin{document}
\title{Jump in fluid properties of inflationary universe to reconcile scalar and tensor spectra}

\author{Hassan Firouzjahi}
\email{firouz-At-ipm.ir}
\author{Mohammad Hossein Namjoo}
\email{mh.namjoo-AT-ipm.ir}

\affiliation{School of Astronomy, Institute for Research in
Fundamental Sciences (IPM),
P.~O.~Box 19395-5531,
Tehran, Iran}

\begin{abstract}

The recent detection of the primordial gravitational waves from the BICEP2 observation seems to
be in tension  with the upper bound on the amplitude of tensor perturbations from the PLANCK data. 
We consider a phenomenological model of inflation in which  the microscopical properties of the inflationary  fluid such as  the equation of state $w$ or the sound speed $c_s$  jump in a sharp manner. We show that  the amplitude of the scalar perturbations  is controlled by a non-trivial combination of $w$ and $c_s$ before and after the phase transition  while the tensor perturbations remains nearly intact.   With an appropriate  choice of the fluid parameters $w$ and $c_s$ one can suppress the scalar perturbation power spectrum on large scales to accommodate a large tensor amplitude with $r=0.2$ as observed by BICEP2 observation.  
 
\vspace{0.3cm}

\end{abstract}

\maketitle

\section{Introduction}
The  BICEP2 observation has reported a detection of B-mode polarization in Cosmic Microwave Background (CMB) on $\ell \sim 100$ \cite{Ade:2014xna} which implies the  detection of the primordial gravitational waves with the  tensor-to-scalar ratio  $r= 0.2^{\, +0.07}_{\, -0.05}$. This high value of tensor amplitude brings friction with the PLANCK date \cite{Ade:2013uln} in which it is found $r< 0.11$ at 95 \% C.L. This is because a high value of $r$ implies a large temperature power spectrum from the sum of the scalar and the tensor perturbations at low $\ell$ multipoles which is not observed.  

The  BICEP2 collaboration already presented a possible resolution to this conflict by allowing
the curvature perturbation spectral index $n_s$ to run with $\alpha_s = d \ln n_s/d \ln k \simeq -0.02$.
However, this large value of $\alpha_s$ is not easy to achieve in simple models of slow-roll inflation in which $\alpha_s \sim (n_s -1 )^2$ so with $n_s \simeq 0.96$ one typically obtains $\alpha_s \sim 10^{-4}$.
As a possible  proposal to remedy this conflict it was argued in \cite{Contaldi:2014zua}, 
\cite{Miranda:2014wga}
that a suppression of curvature perturbation power spectrum $\calP_\calR$ on low multipoles can help to keep the total tensor+ scalar contributions in temperature power spectrum consistent with the PLANCK data, see also \cite{Hazra:2014jka, Abazajian:2014tqa, Smith:2014kka}  for similar line of thought.   The mechanism employed in \cite{Contaldi:2014zua, Miranda:2014wga}, following their earlier works  \cite{Miranda:2012rm, Miranda:2013wxa, Contaldi:2003zv, Park:2012rh}, are based on the scalar field dynamics in which a jump either in the slow-roll parameter $\epsilon \equiv -\dot H/H^2$ or in sound speed $c_s$
is induced to  reduce $\calP_\calR$  on large scales. To see how this works, note that $\calP_\calR$ is given by
$
\calP_\calR = \frac{H^2}{8 \pi^2 M_P^2\epsilon c_s}
$
in which $H$ is the Hubble expansion rate during inflation and  $M_P$ is the reduced Planck mass.
As a result,  one can change  either $c_s$ or   $\epsilon$ to lower $\calP_\calR$ on large scales. 
The idea of producing local features during inflation, like the above two mentioned mechanisms,  
have been studied  vastly in the literature for various purposes, for an incomplete list see  
\cite{Starobinsky:1992ts, Leach:2001zf, Adams:2001vc, Gong:2005jr, Chen:2006xjb, Chen:2008wn, Joy:2007na, Hotchkiss:2009pj,  Abolhasani:2010kn,
Arroja:2011yu, Adshead:2011jq,   Achucarro:2012fd, Cremonini:2010ua, Avgoustidis:2012yc,Romano:2008rr, Ashoorioon:2006wc, Battefeld:2010rf, Firouzjahi:2010ga,  Bean:2008na, Battefeld:2010vr, Emery:2012sm, Bartolo:2013exa, Nakashima:2010sa, Liu:2011cw}


In this paper we present a model of single fluid inflation in which the fluid's microscopical  properties such as the equation of state $w$ or the sound speed $c_s$ undergo a sharp jump. In the spirit this idea is similar to the proposal  employed in \cite{Contaldi:2014zua} and \cite{Miranda:2014wga}. However, we do not restrict ourselves to scalar field theory. Working with the fluid description of inflation will help us to engineer the required jump in fluid's microscopical properties without entering into technicalities associated with the scalar field dynamics. Therefore the results obtained here, based on the general context of fluid inflation, can be employed for the broad class of inflation from a single degree of freedom, in which the scalar field theory is the prime example. 
In addition, we present a careful matching condition on the surface of fluid's phase transition. For a sharp phase transition, after   performing the proper matching conditions,   our analysis shows that some con-trivial combinations of $c_s$ and $w$ controls the final amplitude of curvature perturbation power spectrum. This should be compared from
the usual expectation that  it is the combination $\epsilon c_s$, as appearing in  $\calP_\calR$, which has to jump. This is true for mild phase transition but for a sharp or nearly sharp transition there are additional non-trivial combinations of $\cs$ and $w$ which controls the final amplitude of 
$\calP_\calR$ as we shall see below.

Having this said, we emphasis that our model is a phenomenological one. We do not provide a Lagrangian mechanism  for the fluid. For the simple case of fluid inflation based on a barotropic fluid a Lagrangian formalism is presented in \cite{Chen:2013kta}. In principle a similar Lagrangian formalism can be considered for the general case in which the fluid may be non-barotropic.  Also note that we consider a single fluid with no entropy perturbations. As a result, the curvature perturbations on super-horizon scales remain frozen as we shall verify explicitly. 
 
Note that we need the jump to be extended for one or two e-folds. The first reason is that while we want to reduce $\calP_\calR $ on low multipoles, it  should stabilize to its well-measured value for 
$\ell \gtrsim 100$.  Secondly, a very sharp phase transition will induce dangerous spiky local-type non-Gaussianities and unwanted oscillations in $\calP_\calR$ which may not be consistent with the PLANCK date \cite{Ade:2013uln}. As a result, in our numerical results below we consider the limit in which the duration of phase transition takes one or two e-folds.

\section{The Setup}

In this Section we present our setup. As explained above, we consider a model of inflation based on a single fluid. In addition, we assume that the microscopical properties of the fluid $w$ and $ c_s$
undergo a rapid change from $(w_1, \cs_1)$ during the first phase of inflation to 
$(w_2, \cs_2)$ for the second stage of inflation.  In order to get theoretical insight how the jump in $(w, \cs)$ can help to reduce $\calP_\calR$ we consider the idealistic limit in which the phase transition happens instantly with no time gap.
In this limit we can calculate the final power spectrum analytically and see how the model parameters 
control the result. However, as we mentioned before, to be realistic we need the phase transition to take 
one or two e-folds in order to prevent generating large non-Gaussianities and  unwanted oscillations superimposed on $\calP_\calR$ on small scales.  While our analytical results are for a sharp phase transitions, but  we present the numerical results  for the realistic  case in which transition takes one or two e-folds. 

We assume that in each phase $w$ and $ c_s$ are constant and are  independent free parameters. Only for barotropic fluids one can simply relate $c_s$ and $w$ such as  $c_s^2=w$.  Therefore, in our discussions below, each fluid is labeled  by its parameters $(w_i, \cs_i)$  which are determined from its thermodynamical/microscopical properties.

We do not provide a specific dynamical mechanism for the jumps in $w$ and $\cs$. However,  in principle, one can engineer this  effect by coupling the inflaton fluid to additional fluid or field. For example,  consider the model in which the inflaton field is coupled to a heavy waterfall field. During the first stage of inflation the waterfall is very heavy so we are only dealing with a single field model. Once the inflaton field reaches a critical value the waterfall becomes tachyonic and rapidly rolls to its global minimum. The back-reactions of the waterfall induces a new mass term for inflaton field and will affect its trajectory. As long as the waterfall is heavy and the waterfall phase transition is sharp, one can effectively consider the system as a single field model with the effects of the waterfall instability to cause a sudden change in slow roll parameters.

\subsection{The Background Dynamics}

With these discussions in mind, let us proceed with our analysis.  
The background is  a flat FLRW universe with the metric 
\ba
ds^2 =  -dt^2+ a(t)^2 d \bfx^2=a^2(\eta)(-d\eta^2+d\bfx^2) \,,
\ea
in which $\eta$, defined by $d\eta = dt/a(t)$, is the conformal time.

Denoting the energy density and the pressure of the fluid by $\rho$ and $P$ respectively, 
the equation of state is $w= P/\rho$.   We assume inflation has two stages separated at $\eta = \eta_{*}$. For the period $\eta < \eta_{*}$ our parameters are  $w=w_1, \cs= \cs_1$ while for
$ \eta_{*}< \eta < \eta_e$ they are  $w= w_2, \cs= \cs_2$ in which $\eta_e$ represents the time of end of inflation.  In order to support inflation we assume $-1< w_i < -1/3$, while for the slow-roll model we may further assume that $1+ w_i \rightarrow 0$.

Using the energy conservation equation  the evolution of energy density  is given by
\ba
\rho= \rho_* \left(\dfrac{a}{a_*} \right)^{-3(1+w)}\,.
\ea 
in which $\rho_*$ represents the value of $\rho$ at the  time of phase transition (with similar definition for other quantities with the subscript $*$).  Note that the above equation is valid for each phase so we have removed the index $i$. 

In addition,  the Friedmann equation  is 
\ba
3 \mPl^2 \calH^2= a^2 \rho\,,  
\ea
where ${\cal H}={a'}/{a} $ is the conformal Hubble parameter in which
a prime represents the derivative with respect to conformal time $\eta$.
One can easily integrate the Friedmann equation  to  get
\begin{eqnarray}
\label{a-eq}
a_i(\eta) = a_*\Bigl(\frac{{\cal H}_*}{\beta}(\eta-\eta_*)+1\Bigr)^{\beta_i}\,,
\end{eqnarray}
where $a_i(\eta)$ means the value of $a(\eta)$ during each phase $i=1, 2$  and we have defined 
the parameter $\beta_i$ via
\begin{eqnarray}
\label{beta}
\quad \beta_i\equiv\frac{2}{3w_i+1}\,.
\end{eqnarray}
Taking the conformal time derivative from Eq.~(\ref{a-eq}) we get
\ba
\calH_i=\frac{\calH_*}
{1+\displaystyle\frac{\calH_*}{\beta_i}\,( \eta- \eta_* )}\,.
\ea
One can check that both $a(\eta)$ and ${\cal H}$ are continuous at the time 
of phase transition $\eta=\eta_*$ when $w$ and $\cs$ undergo a sudden change. 

\subsection{The Perturbations}
Now we study the scalar and the tensor perturbations in this setup.  The analysis 
of the scalar perturbations are similar to the analysis in  \cite{Namjoo:2012xs} and here we outline the main  results.

\subsubsection{Scalar Perturbations}
\label{scalar-pert}

The equation of motion for the curvature perturbations on comoving surface $\calR$ for a 
fluid with the known equation  of state parameter $w$ and sound speed
$c_s$ in  the Fourier space is given by
\ba
\label{R-eq}
{\cal R}_\bmk'' + \dfrac{(z^2)'}{z^2} {\cal R}'_\bmk 
+c_s^2 k^2 {\cal R}_\bmk=0  \, ,
\ea 
where
\ba
z \equiv a(\eta) \mPl \sqrt{3(1+w)}/c_s\,.
\ea 
Note that $c_s$ is defined as $\delta P_c = c_s^2 \delta \rho_c$ 
in which the subscript c indicates that the corresponding quantities 
are calculated on the comoving hypersurface . 

For constant  values of $w$ and $c_s$ one can easily solve Eq.~(\ref{R-eq}) in each phase. 
However, one has to perform the matching conditions at the time of transition $\eta = \eta_*$ to match the  outgoing solutions to the incoming solutions \cite{Deruelle:1995kd}.
The first matching condition is that  the curvature perturbation to be continues 
\ba
\label{R-bc1}
[{\cal R}_\bmk]_-^+ =0   \, ,
\ea
where  $[X]_-^+$ denotes the difference in  $X$ 
after and before the transition: $[X]_-^+ = X(\eta_+) - X(\eta_-)$.
Geometrically, the continuity of ${\cal R}$ is interpreted as the
continuity of the extrinsic and the intrinsic curvatures on the 
three-dimensional spatial hyper-surfaces located at 
$\eta=\eta_{*}$.

To find the  matching condition for the time derivative
 of ${\cal R}$ we  note that Eq.~(\ref{R-eq}) can be rewritten as
\ba
\dfrac{d}{d\eta} \left(\frac{a^2}{c_s^2} (1+w) {\cal R}'_\bmk \right)
+a^2 (1+w)  k^2 {\cal R}_\bmk=0  \, .
\ea
Integrating the above equation in a small time interval around the surface of the phase
transition, the last term above vanishes and we obtain the second  matching condition as follows
\ba
\label{matchR2}
\left[ \dfrac{1+w}{c_s^2} {\cal R}_\bmk'\right]_\pm =0 \, .
\ea
Note the non-trivial combination $(1+w)/c_s^2$ which appears in this matching condition.
This will play important roles when we calculate the final power spectrum after performing the matching conditions.

One can  easily  solve the equation of motion for ${\cal R}_\bmk$.
For constant values of $w$ and $c_s$, we have $z'/z=a'/a=\calH$. As a result Eq.~(\ref{R-eq}) simplifies to
\begin{eqnarray}
\left[\frac{d^2}{dx^2}+\frac{2\beta}{x}\frac{d}{dx}+1\right]{\cal R}_\bmk=0  \, ,
\end{eqnarray}
where 
\begin{eqnarray}
\label{xdef}
x\equiv- c_s \, k (\eta-\eta_*+\beta\calH_*^{-1})\,.
\end{eqnarray}
The solution is 
\ba
\label{R-sol}
{\cal R}_\bmk
= x^{\nu} \left[ C_1  H^{(1)}_\nu\left(x\right) 
+D_1 H^{(2)}_\nu\left(x\right) \right] \,;
\quad \nu\equiv \frac{1}{2}-\beta=\frac{3(w-1)}{2(3w+1)}  \,,
\ea
in which $C_1$ and $D_1$ are constants of integration
while $H^{(1)}_\nu(x)$ and $H^{(2)}_\nu(x)$ are the Hankel functions of 
the first and second kinds, respectively. 
Note that for slow-roll inflation with $w \simeq -1$, we have $ \nu \simeq 3/2$. 

For the modes deep inside the horizon the solution should approach 
the Minkowski positive frequency mode function so for  $\eta \to -\infty$ we require 
\ba
{\cal R}_\bmk \to  \dfrac{e^{-i c_s k \eta}}{z(\eta) \sqrt{2 c_s k}}
\quad 
\mathrm{for}  \ \eta\to -\infty\,.
\ea
Imposing this initial condition on  $\calR_\bfk$ and 
using the asymptotic form of the Hankel function we find 
$D_1=0$ and
\ba
\label{R1}
{\cal R}_\bmk(\eta)=C_1\,
x_1(\eta)^{\nu_1} H^{(1)}_{\nu_1}(x_1(\eta))  \, ;
\quad \eta<\eta_{*}\,,
\ea
where 
\begin{eqnarray}
x_1&\equiv& - c_{s1}\, k(\eta-\eta_{*}+\beta_1\calH_{*}^{-1})\,;
\quad \nu_1=\frac{1}{2}-\beta_1=\frac{3(w_1-1)}{2(3w_1+1)}\,,
\end{eqnarray}
and 
\ba 
C_1 \equiv  \dfrac{-1}{2 \mPl \, a_*}
\left( \dfrac{ i \pi c_{s1} e^{i \pi \nu_1}}{3 k (1+w_1)} \right)^{1/2}
 x_{1*}^{1/2-\nu_1} \, .
\ea
Not that $x_{1*} = x_1(\eta_*)$ and so on.

We would like to calculate  the curvature perturbation ${\cal R}_\bmk$ at
 the end of inflation $\eta_e\simeq 0$.
The curvature perturbation power spectrum ${\cal P}_{{\calR}}$ is given by
\ba
\label{R-powr}
\langle  {\cal R}_{ \bmk} {\cal R}_{ \bmk'}\rangle 
\equiv (2\pi)^{3} P_{{\cal R}}(k)~\delta^3(\bmk+\bmk')\,,
\quad
{\cal P}_{\cal R}\equiv \frac{k^{3}}{2 \pi^{2}}P_{\cal R}(k)\,;
\quad P_{\calR}(k)=|\calR_\bmk|^2\,,
\ea

For the modes which leave the sound horizon during the first stage of inflation the power spectrum at the time $\eta = \eta_*$ in the slow-roll limit  $(\epsilon \ll 1) $ is given by 
\ba
{\cal P}_{{\calR}}(\eta_{*}) 
\simeq \dfrac{H_{*}^2}{8 \pi^2 \,\mPl^2  c_{s1} \, \epsilon_1}\,,
\ea 
in which the slow-roll parameter $\epsilon$ is  
\begin{eqnarray}
\epsilon \equiv - \frac{\dot H}{H^2}=\frac{3}{2}(1+w)\,.
\label{epsilondef}
\end{eqnarray}
Since we work with a single adiabatic fluid the curvature perturbation is conserved
if this mode  remains super-horizon until the end of inflation, and we have
\begin{eqnarray}
{\cal P}_{{\calR}}(\eta_{e})= {\cal P}_{{\calR}}(\eta_{\eta_*})
\simeq \dfrac{H_{*}^2}{8 \pi^2 \,\mPl^2  c_{s1} \, \epsilon_1}\,.
\label{PR1}
\end{eqnarray}

For the second stage of inflation we have
\ba
\label{R2}
{\cal R}_\bmk(\eta) 
&=& C_2 \, x_2^{\nu_2} H^{(1)}_{\nu_2}\left(x_2(\eta) \right)
+ D_2 \, x_2^{\nu_2} H^{(2)}_{\nu_2}\left(x_2(\eta) \right) \,;
\quad \eta_{*}<\eta<\eta_{e}\,,
\ea
where $x_2(\eta)$ is defined in accordance with
the general definition~(\ref{xdef}),
\begin{eqnarray}
x_2&\equiv& -c_{s2}\, k(\eta-\eta_{*}+\beta_2\calH_{*}^{-1})\,;
\quad \nu_2=\frac{1}{2}-\beta_2=\frac{3(w_2-1)}{2(3w_2+1)}\,,
\end{eqnarray}

Imposing the matching condition for $\calR$ and $\calR'$ at $\eta= \eta_*$, we 
can calculate $C_2$ and $D_2$ in terms of $C_1$.  Following \cite{Namjoo:2012xs}
we obtain 
\ba
\label{exact1}
C_2&=& -\dfrac{\pi \, x_{1*}^{\nu_1}}{4 i \, x_{2*}^{\nu_2-1}}
C_1 \left[H^{(1)}_{\nu_1} \left(x_{1*} \right) 
H^{(2)}_{\nu_2-1}\left(x_{2*}\right)
- f H^{(1)}_{\nu_1-1} (x_{1*}) 
H^{(2)}_{\nu_2} \left(x_{2*}\right) \right]\,,
\\ 
\label{exact2}
D_2 &=&\dfrac{\pi \, x_{1*}^{\nu_1}}{4 i \, x_{2*}^{\nu_2-1}}
 C_1 \left[H^{(1)}_{\nu_1} (x_{1*}) 
H^{(1)}_{\nu_2-1}(x_{2*})- f 
H^{(1)}_{\nu_1-1} (x_{1*}) H^{(1)}_{\nu_2}
 (x_{2*}) \right]\,,
\ea
where we have defined 
\ba
f \equiv  \dfrac{(1+w_1)}{(1+w_2)}  
\frac{c_{s2}}{c_{s1}}  = \frac{\epsilon_1 \cs_2}{\epsilon_2 \cs_1} \, .
\ea

As usual, we are interested in modes which are super-horizon at
the end of inflation $x_2(\eta_e) \ll 1$ for $\eta_e \rightarrow 0$.
Using the small argument limit of  of the Hankel function, we obtain
\ba
 \label{R-end}
{\cal R}_\bmk(\eta_e \to 0) \simeq - \dfrac{i\,  2^{\nu_2}}{\pi}\,
 \Gamma(\nu_2)   (C_2-D_2)\,.
\ea
As in \cite{Namjoo:2012xs} we define the transfer function $T_\calR$ for the curvature perturbation power spectrum  as
\ba
{\cal P}_{\cal R}(\eta_e) = T_\calR \,{\cal P}_{{\cal R}_1}(\eta_e) \,,
\ea
where ${\cal P}_{{\calR}_1}(\eta_e)$ is the power spectrum at 
the end of inflation in the absence of any change in $w$ and $\cs$, i.e. when  $w=w_1$ and 
$\cs =\cs_1$ throughout the inflationary stage,
as calculated in Eq. (\ref{PR1}).
Assuming $\nu_1 \simeq \nu_2 \simeq 3/2$ we obtain
\ba
\label{T-def}
T_\calR \simeq \dfrac{\vert C_2 - D_2 \vert^2}{\vert C_1 \vert^2}  \, .
\ea
In this view  any non-trivial effect due to change in $w$ and $\cs$ is captured by 
the  transfer function $T_\calR$.

Using the results  for $C_2$ and $D_2$ given in Eqs. (\ref{exact1}) and (\ref{exact2})
we obtain
\ba
\label{TR}
T_\calR = \left(  \frac{\pi x_{1*}^{\nu_1}}{2 x_{2*}^{\nu_2 -1}} \right)^2 \left | H_{\nu_1}^{(1)} (x_{1*})  J_{\nu_2-1} (x_{2*}) - f  H_{\nu_1-1}^{(1)} (x_{1*})  J_{\nu_2} (x_{2*})
\right |^2  \, .
\ea
Note that \ba
\label{x12-def}
x_1(\eta_{*}) = -\frac{\cs_1 k \beta_1}{\calH_{*}} \quad , \quad 
x_2(\eta_{*}) = -\frac{\cs_2 k \beta_2}{\calH_{*}} \, .
\ea

The discussions above were general with no slow-roll assumptions. To get better insights into the results, let us consider the simple and more realistic case in which $w_i \simeq -1$ corresponding to  $\epsilon_i \simeq 0$. In this limit, $\beta_i$ and $\nu_i$ are nearly insensitive to  
the change in $w$ or $\epsilon$ and one practically takes $\beta_i = -1$ and $\nu_i=3/2$. In this limit
$x_{1*} = \cs_1 k_*$ and  $x_{2*} = \cs_2 k_*=  \frac{\cs_2}{\cs_1}k_*$ in which we have defined  
\ba
k_* \equiv \frac{ k}{\calH_*} \, .
\ea
Note that $k_*$ represents the mode which leaves the sound horizon at the time $\eta_*$ in the absence of phase transition, i.e. when $w_1=w_2$ and $\cs_1 = \cs_2$. In the slow-roll approximations the transfer function simplifies to 
\ba
\label{TR-slow}
T_\calR(k) = \left(  \frac{\pi^2 k_*^2 \cs_1^{3}}{4 \cs_2} \right) \left | H_{\frac{3}{2}}^{(1)} ( \cs_1 k_*)  J_{\frac{1}{2}} (\frac{\cs_2}{\cs_1}k_*) - f  H_{\frac{1}{2}}^{(1)} (\cs_1 k_*)  J_{\frac{3}{2}} (\frac{\cs_2}{\cs_1}k_*)
\right |^2  \,  \quad \quad \mathrm{ (slow-roll )}
\ea

The interesting effect is that   depending on the ratio $\cs_2/\cs_1$ the inverse comoving sound horizon,  $\calH/c_s$, shows non-trivial behaviors.  First assume that $\cs_2 < \cs_1$. 
In this limit the structure of the comoving sound horizon  is similar to conventional models of inflation in which  $\calH/\cs$ is an increasing function in both periods of inflation. As a result, a mode which is outside the sound horizon during the first period of inflation, corresponding to 
$x_{1*} = \cs_1 k_* <1$, will also be outside the sound horizon during  the second stage of inflation with  $x_{2*}= \cs_2 k_*<1$. Using the small argument limit of the Bessel functions, one can easily check from Eq. (\ref{TR-slow}) that $T_\calR \rightarrow 1$. This is expected since we work with a single adiabatic fluid so $\calR$ is blind to changes in $w$ and $c_s$ on super-horizon scales. 
In addition, there are modes which are 
inside the sound horizon during the first stage of inflation, $ \cs_1 k_* >1$,  but leaves the sound horizon during the second stage, $ \cs_2 k_* <1$.  As expected, $T_\calR(k)$ shows non-trivial behaviors for these modes.

\begin{figure}
\includegraphics[width =  3in ]{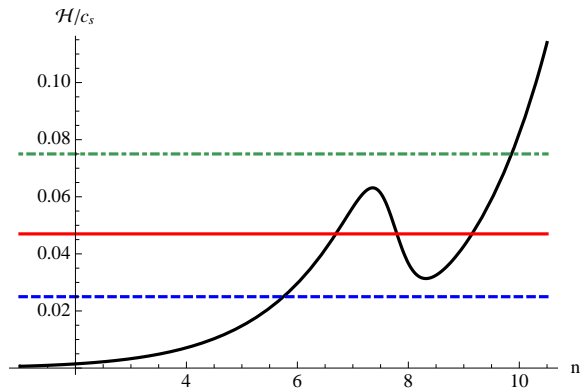}
\caption{ The plot of the inverse comoving sound horizon $\calH/c_s$ 
as a function of the number of e-folds $n$ for the case II in which $\cs_2 > \cs_1$. There are three different distinct behaviors for the modes. The lower dashed line represents modes which are outside the sound horizon  during both stages of inflation. The  middle solid line represents modes which leave the sound horizon during the first stage of inflation, re-enter the sound horizon during the second stage of inflation and leave the sound horizon again during the second period of inflation.
The  the upper dashed-dotted line represents modes which are inside the sound horizon during both periods of inflation. Here  $c_{s1}=0.2, c_{s2}=1.0, w_1=-0.96, w_2=-0.996$. }
\label{comov-H} 
\end{figure}

Now consider the case in which $\cs_2 >\cs_1$ so $x_{2*} > x_{1*}$. In this limit $\calH/\cs$ decreases during the second 
stage of inflation briefly before increasing again. A plot of the behavior of $\calH/\cs$ is given
in Fig. \ref{comov-H}. Now there are three possibilities for whether  a mode is super-horizon or sub-horizon.
(a): There are modes which are outside the sound horizon in both periods of inflation, corresponding to the case where $x_{2*}  <1$. In this limit, one can easily check that $T_\calR \rightarrow 1$
as explained before. (b): There are modes which are outside the sound horizon during the first stage
of inflation, but re-enter the sound horizon during the second stage of inflation and after some time
leave the sound-horizon again. This is a new behavior which does not exist in conventional models of inflation in which  $\calH/\cs$ is always increasing.
(c): This corresponds to case in which the mode is inside the sound horizon during both stages of inflation  with $ x_{1*} >1$.  

We present the analytical and  numerical plots of $\calP_\calR $ in next Section.

\subsubsection{The Tensor Perturbations}

Now we study tensor perturbations in this background.

The equation of motion for the tensor perturbation $h_{ij}$ subject to $\partial_j h_{ij} = h_{ii}=0$ (sum over the repeated indices) is given by
\ba
\label{tensor-eq}
D_k'' + 2 \calH D_k' + k^2 D_k =0
\ea
where we have used the decomposition $h_{ij} \sim e_{ij} D_k e^{i \bfk \cdot \bfx}$
in Fourier space in which $e_{ij}$ represents the two polarizations of the tensor perturbations. 

We note that  the equation Eq. (\ref{tensor-eq}) for tensor perturbation has the same form
as the scalar perturbations Eq. (\ref{R-eq}) with the replacement $z= a$ and $c_s=1$. The latter 
condition is originated from the fact that the tensor perturbations propagate with the speed of massless
particles which is set to unity.  In addition, note that the parameter $w$ does not enter into tensor
perturbations directly, it only enters indirectly through the background expansion $\calH$. 

The solution for the tensor perturbations has the same form as Eq. ( \ref{R-sol}) for the first phase and
Eq. ( \ref{R2}) for the second phase. After imposing the matching conditions and defining the transfer function for the tensor perturbations $T_T$ similar to scalar perturbations, we obtain
\ba
\label{TT}
T_T = \left(  \frac{\pi y_{1*}^{\nu_1}}{2 y_{2*}^{\nu_2 -1}} \right)^2 \left | H_{\nu_1}^{(1)} (y_{1*})  J_{\nu_2-1} (y_{2*}) -  H_{\nu_1-1}^{(1)} (y_{1*})  J_{\nu_2} (y_{2*})
\right |^2  \, ,
\ea
in which we have defined $y_i = x_i/\cs_i$. The difference between $x_i$ and $y_i$ is originated from the fact that  the tensor perturbations propagate  with the speed equal to unity and do not see the sound speed $c_s$. In addition,  note that unlike $T_\calR$, in  $T_T$ expression  the factor $f$ did not appear. This is because the factor $f$ is originated from the factor 
$(1+ w)/c_s^2$ in the matching condition Eq. (\ref{matchR2}).

The dependence of $T_T$ to changes in background only come via the factor $\beta_i$ or $\nu_i$
which are not sensitive functions of $w$. Therefore, one expects that  the gravitational wavs are nearly insensitive to the changes in $w$ or $\cs$.
To see this more easily, consider the slow-roll limit in which $\nu_i =3/2$, $\beta_i=-1$ and 
$y_{1*} = y_{2*}$. In this approximation, Eq. ( \ref{TT}) simplifies to
\ba
T_T \simeq  \frac{\pi^2 y_{1*}^{3}}{4 y_{1*}}  \left | H_{\frac{3}{2}}^{(1)} (y_{1*})  J_{\frac{1}{2}} (y_{1*}) -  H_{\frac{1}{2}}^{(1)} (y_{1*})  J_{\frac{3}{2}} (y_{1*})
\right |^2  =1 \, .
\ea
We also checked numerically that $T_T$ has only a mild dependence to the changes in $w$ and $\cs$.

Finally,  let us denote the ratio of the two transfer functions  by $\gamma \equiv T_T/T_\calR$. From the above discussions it is clear that $\gamma(k)$ inherits most of its non-trivial behavior from 
$T_\calR$.   The ratio of the tensor power spectrum to the scalar power spectrum   $r \equiv \calP_T/\calP_\calR$ is related to $\gamma$ via 
\ba
r(k) = \gamma(k) r(k_*) \, .
\ea
We see that a non-trivial scale-dependence in $\gamma(k)$ is translated into a scale-dependent 
of $r(k)$. \\

\begin{figure}
\includegraphics[width =  3in ]{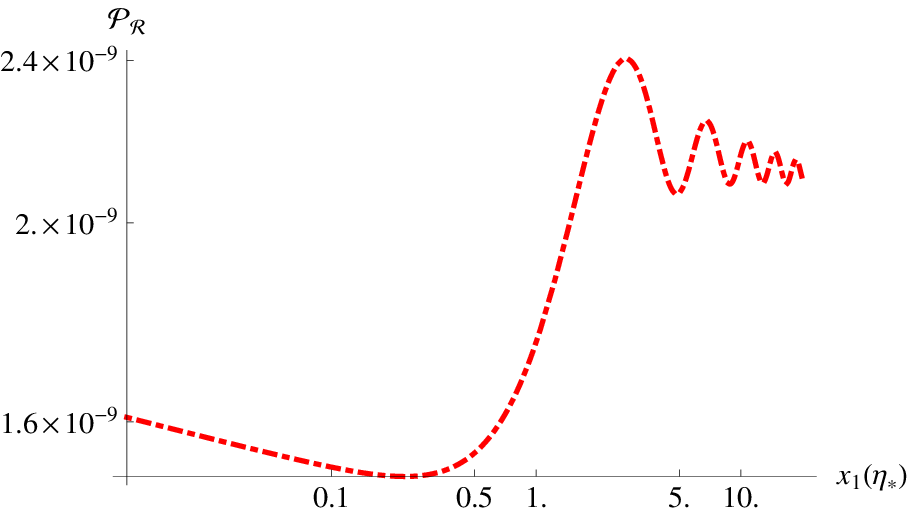}
\hspace{1cm} \vspace{1cm}
\includegraphics[width =  3in ]{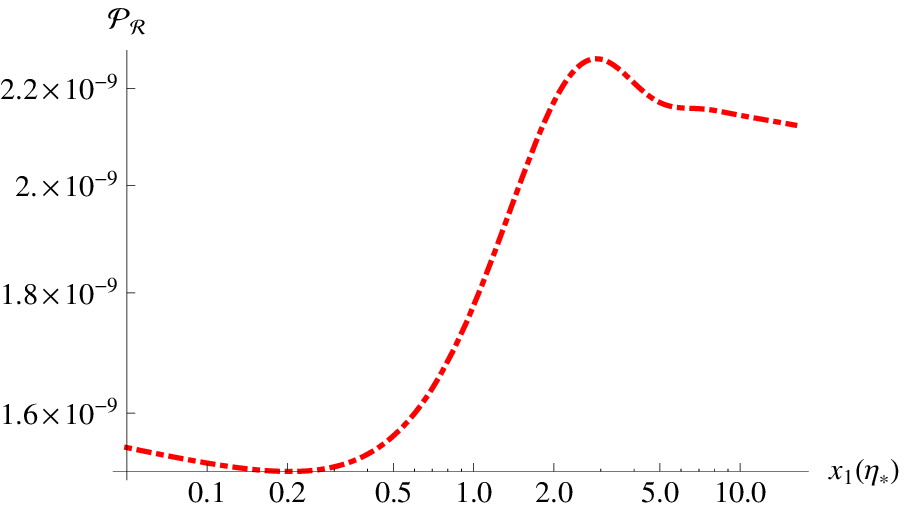}
\vspace{0.cm}
\includegraphics[width =  3in ]{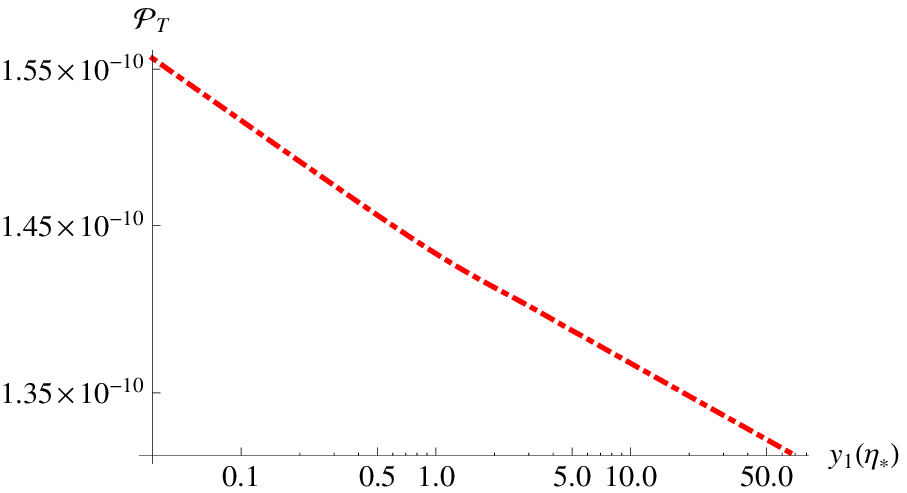}
\caption{ 
 Case I: $c_{s1}=1.0, c_{s2}=0.8, \epsilon_1=0.012, \epsilon_2=0.01$ . The left plot is for an infinitely sharp 
phase transition while the right plot represents the case in which the phase transition takes
1.5 e-folds.  The bottom plot if for $\calP_T$. As we see, the tensor perturbations are nearly insensitive to the phase transition. }
\label{T-radiation1} 
\end{figure}

\section{The Numerical Results}

Here we present our numerical results to see as how a jump in $(w, \cs)$ can reconcile the tension between the PLANCK and the BICEP2 data. As we explained before, this tension can be alleviated if
we consider the situation in which $\calP_\calR$ is reduced for low-$\ell$ multipoles so the total
combination of tensor+scalar power spectra matches the PLANCK observations in this regions. 

We present the plots for both the idealistic case with
arbitrarily sharp phase transition and the realistic case in which the phase transition is mild enough to get rid of unwanted non-Gaussinities and oscillations in $\calP_\calR$.  
In our numerical analysis for the not too sharp phase transition we assume the phase transition takes 1.5 e-folds.

First we consider the case in which $\cs_2 < \cs_1$ (in our numerical plots, this case is known as case I).  As discussed before, in this case $\calH/\cs$ is increasing during both periods of inflation so the comoving sound horizon has the same structure as in conventional models. In Fig. \ref{T-radiation1} we have presented the power spectrum for 
both the infinitely sharp transition and the near mild transition. As described before, assuming that the phase transition is not infinitely sharp will eliminate the unwanted oscillations on small scales.  
For modes which are outside the sound-horizon $\calP_\calR$ is nearly insensitive 
to the change in $w$ and $\cs$ and it follows its original nearly scale-invariant shape with a small red-tilt. However, for modes which are inside the sound horizon during the first stage of inflation the situation is nontrivial.    This corresponds to the case in which $x_{1*} >1$ but
$x_{2*} <1$. Using the small and the large arguments limit of the Hankel functions, one can check
that $T_\calR \sim k^2$. This strong k-dependence is seen near $x_{1*}>1 $ in Fig. \ref{T-radiation1}. Finally  for the modes which are inside the sound horizon in both periods of inflation $\calP_\calR$ reaches nearly a constant value in which the oscillations are damped assuming the phase transition is mild enough.  Physically this makes sense, since very small scales are blind to nearly mild
phase transitions which happened in the past inflationary history. 

\begin{figure}
\includegraphics[width =  3in ]{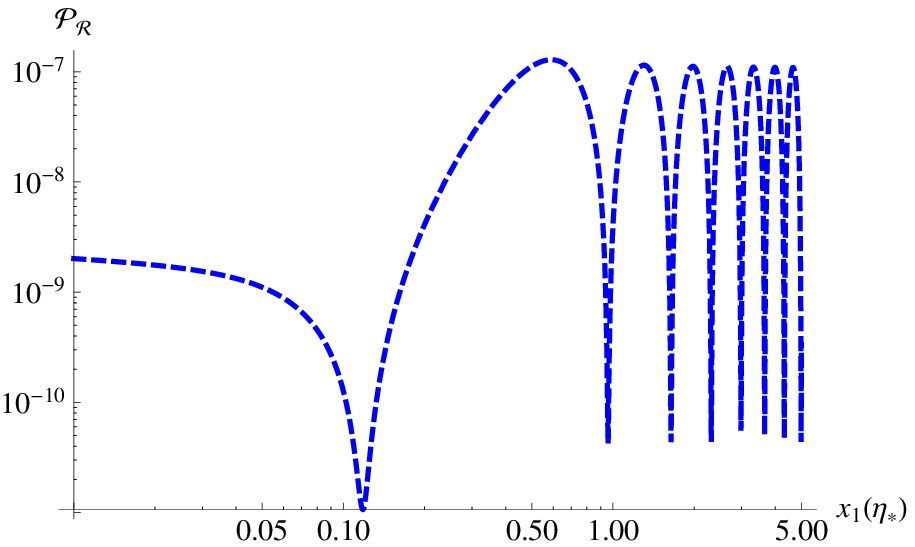}
\hspace{1cm} \vspace{1cm}
\includegraphics[width =  3in ]{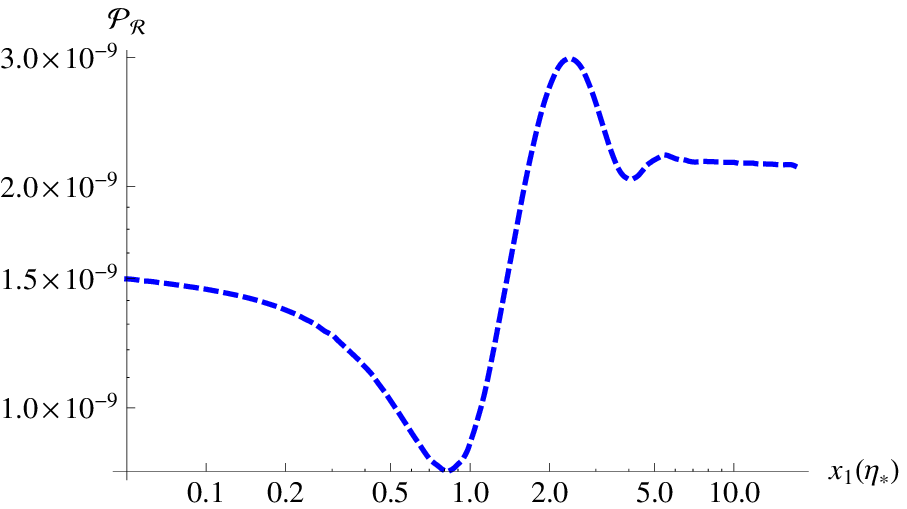}
\includegraphics[width =  3in ]{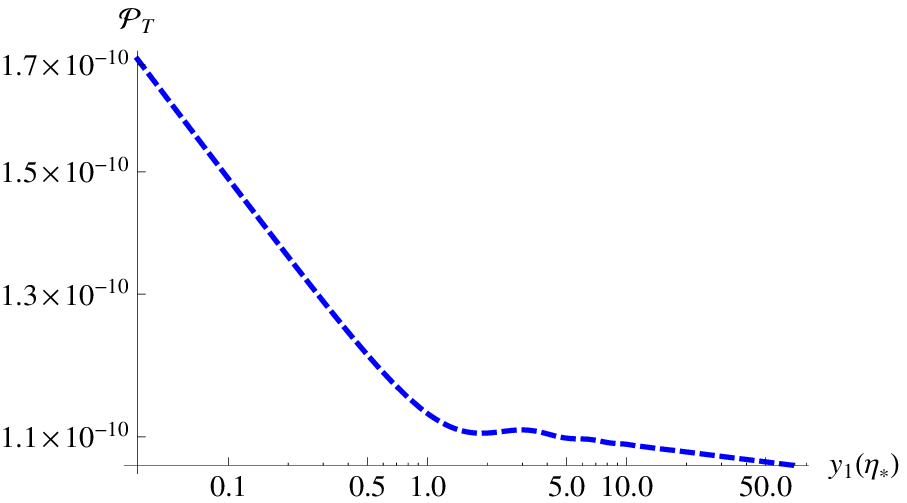}
\caption{ The same plot as in Fig. \ref{T-radiation1} but for
case II: $c_{s1}=0.8, c_{s2}=1.0, \epsilon_1=0.016, \epsilon_2=0.007$.
 Note that the qualitative shape of the jump in $\calP_\calR$ is different than the case in Fig. \ref{T-radiation1}. The bottom plot represents the tensor power spectrum. Interestingly, the tensor perturbations are more affected compared to case I in  Fig. \ref{T-radiation1} since $\epsilon_2$ has decreased significantly in order to maintain $\epsilon_2 \cs_2 < \epsilon_1 \cs_1$. 
 }
\label{T-radiation2} 
\end{figure}

Now we consider the interesting case in which $\cs_2 > \cs_1$ (denoted in numerical plots by case II).  Of course, to enhance the final power spectrum we need $\epsilon_2 < \epsilon_1$ such that $\cs_2 \epsilon_2 < \cs_1 \epsilon_1$.
In Fig. \ref{T-radiation2} we have presented the shape of the final power spectrum for this case. 
As described at the end of subsection  \ref{scalar-pert} and in Fig. \ref{comov-H}
there are three possibilities labeled (a), (b) and (c) for whether the mode is inside the sound horizon or
outside the sound horizon. 
For case (a) the mode is outside the sound horizon during both stages of inflation and 
$T_\calR \rightarrow 1$. The case (b) is non-trivial in which the mode is outside the sound horizon
during the first period of inflation, re-enter the sound horizon during the second stage of inflation
and leave the sound-horizon again. The shape of $T_\calR$ and the final power spectrum are complicated functions which depend non-trivially on $\cs_i$ and $w_i$. In particular, the factor
$f= \epsilon_1 \cs_2/\epsilon_2\cs_1$ plays a nontrivial role.  Note that with $\cs_2 < \cs_1$ and
$\epsilon_1 > \epsilon_2$, $f$ can be a relatively large number. In this case, using the asymptotic limit
of Bessel functions, we get $T_\calR \sim \frac{\cs_2 \epsilon_1^2}{\cs_1 \epsilon_2^2}$. Finally for the case (c) in which the mode is inside the sound horizon in both stages of inflation, we have
$T_\calR \sim \frac{ \epsilon_1^2}{ \epsilon_2^2}$. We have checked that both of these estimations for $T_\calR$ for cases (b) and (c) are in reasonable agreement with the full numerical analysis with a relatively mild phase transition. 
 \\

\begin{figure}
\includegraphics[width =  3.05in ]{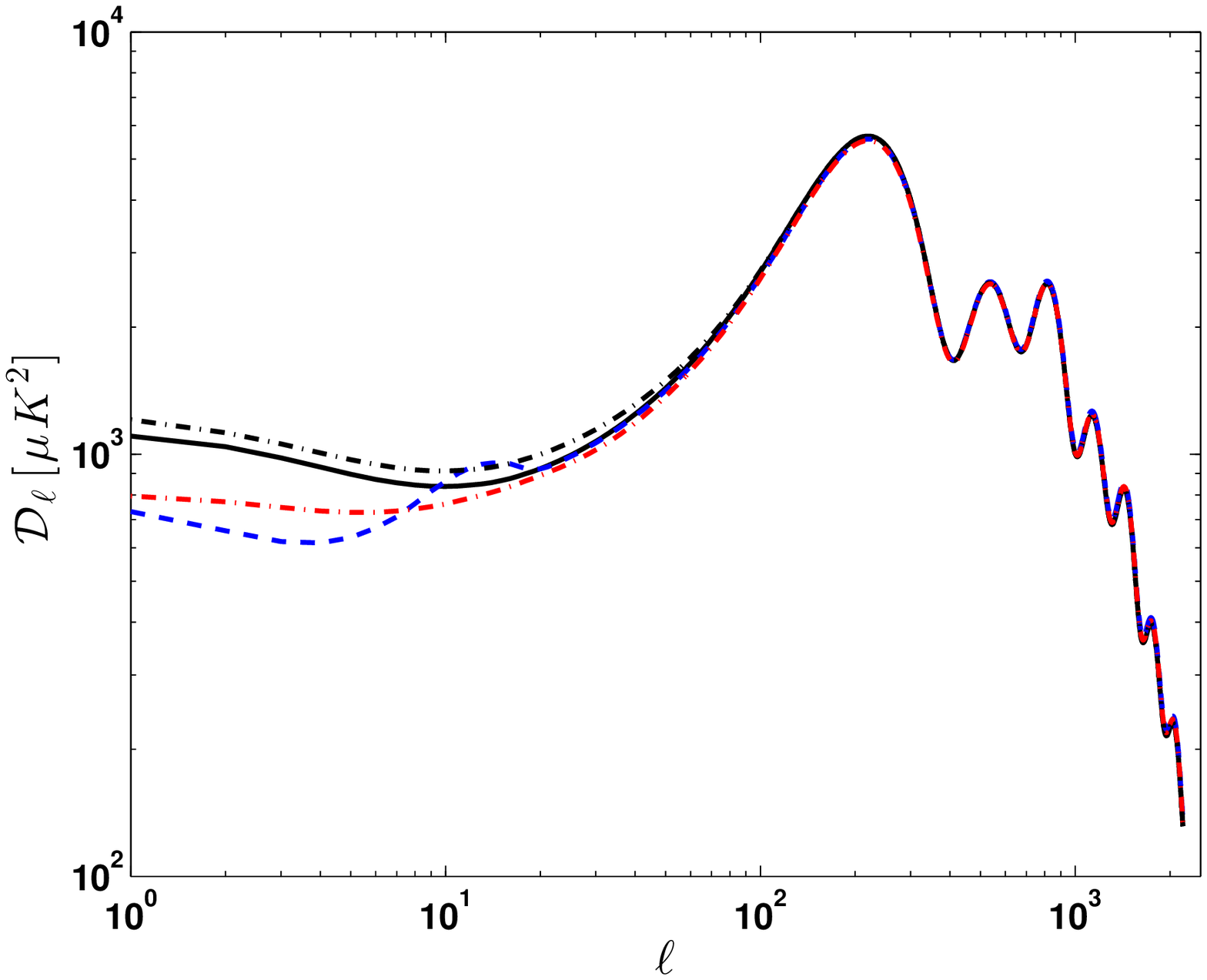}
\hspace{1cm}
\includegraphics[width =  3in ]{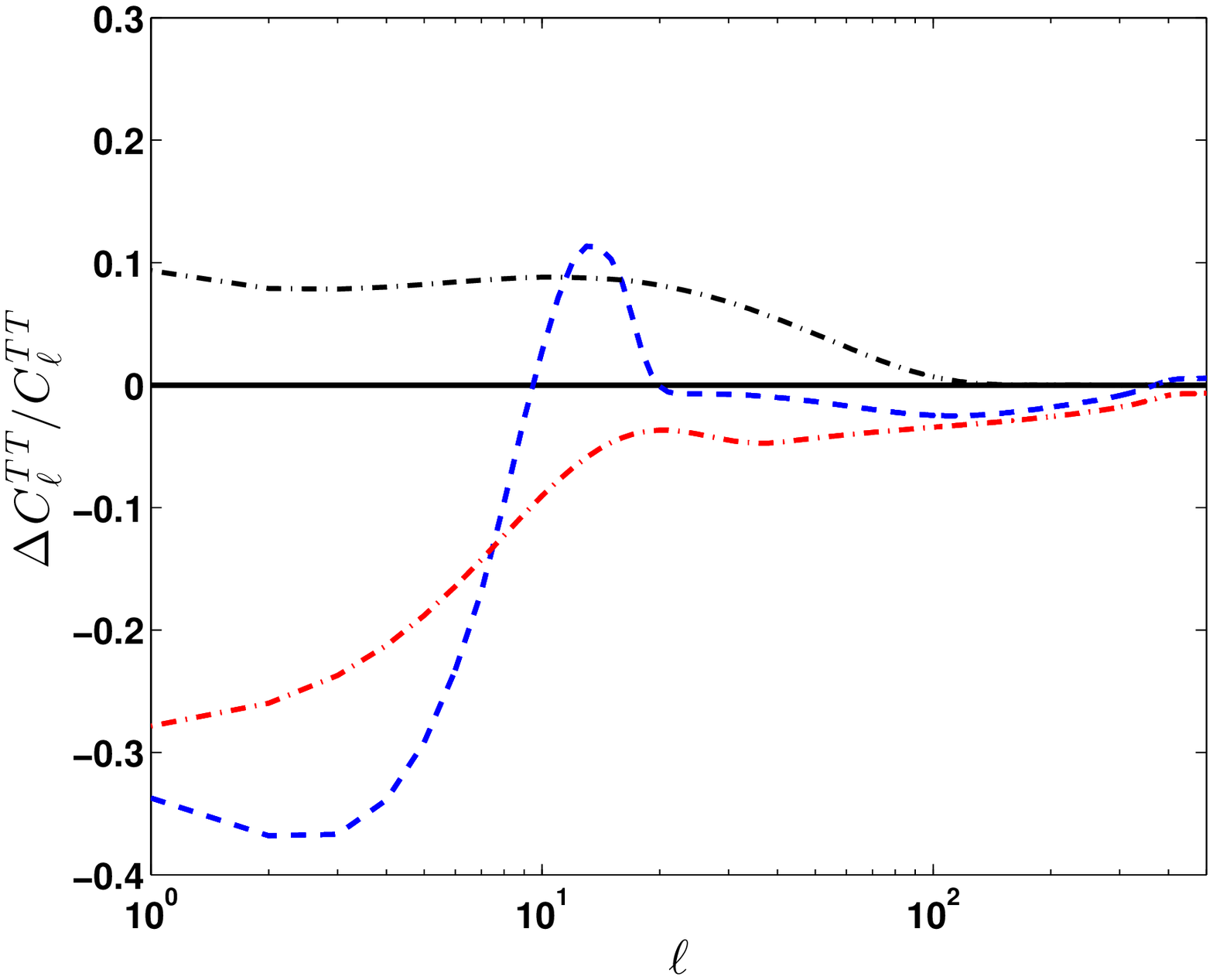}
\caption{Here we present the TT power spectrum in which ${\cal D}_\ell$ is related to $C_\ell$ via
${\cal D}_\ell = \ell (\ell+1) C_\ell/2\pi$. The red dashed-dotted curve and the 
blue dashed curve, as in previous plots, correspond  to case I ($\cs<1$) and II  ($\cs>1$) respectively.
The   solid  black curve represent the  standard $\Lambda$CDM with $r=0$. The upper dotted-dashed-black curve is for  $\Lambda$CDM with $r=0.2$.   }
\label{TT} 
\vspace{0.5cm}
\end{figure}

\begin{figure}
\includegraphics[width =  3.05in ]{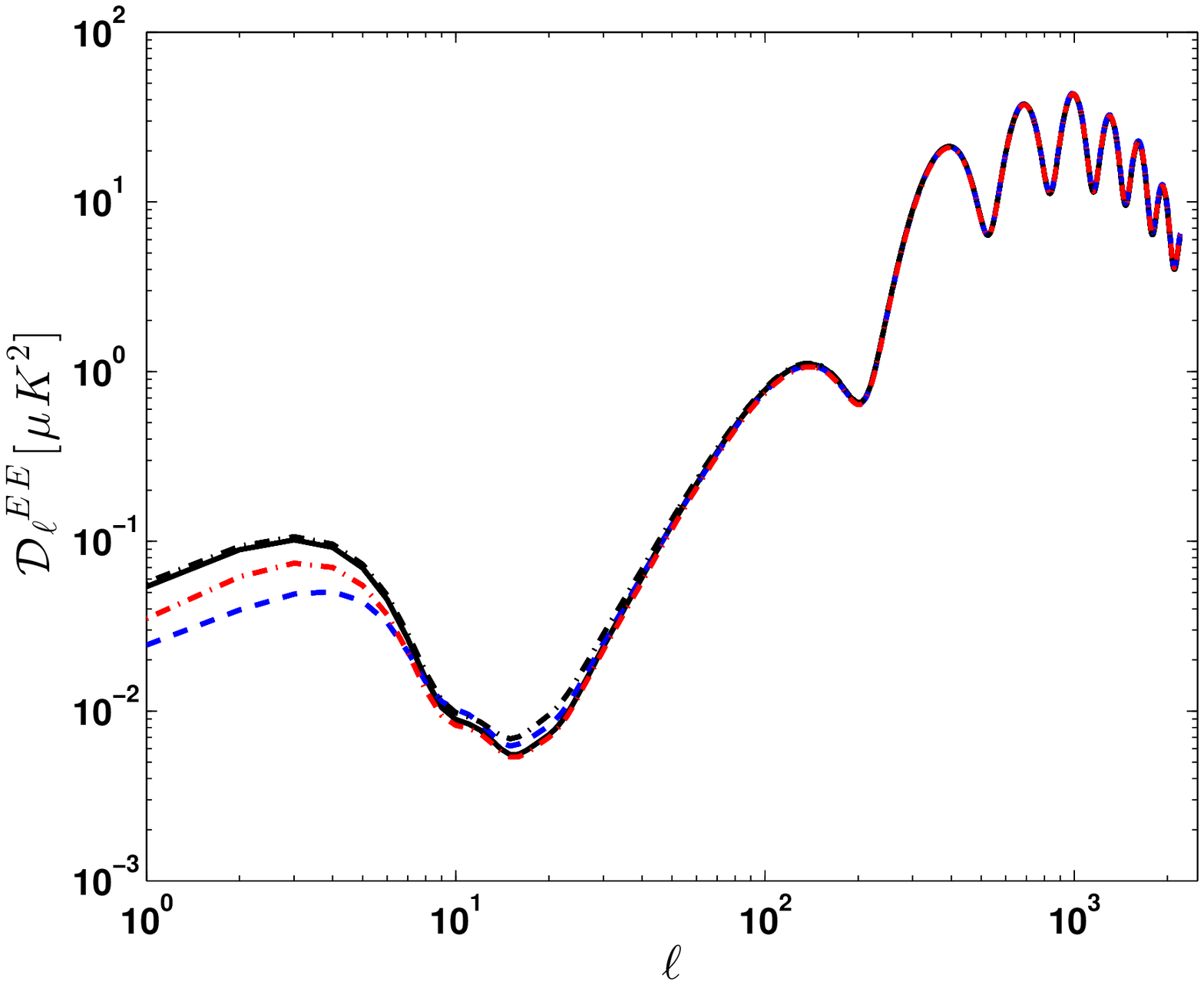}
\hspace{1cm}
\includegraphics[width =  3in ]{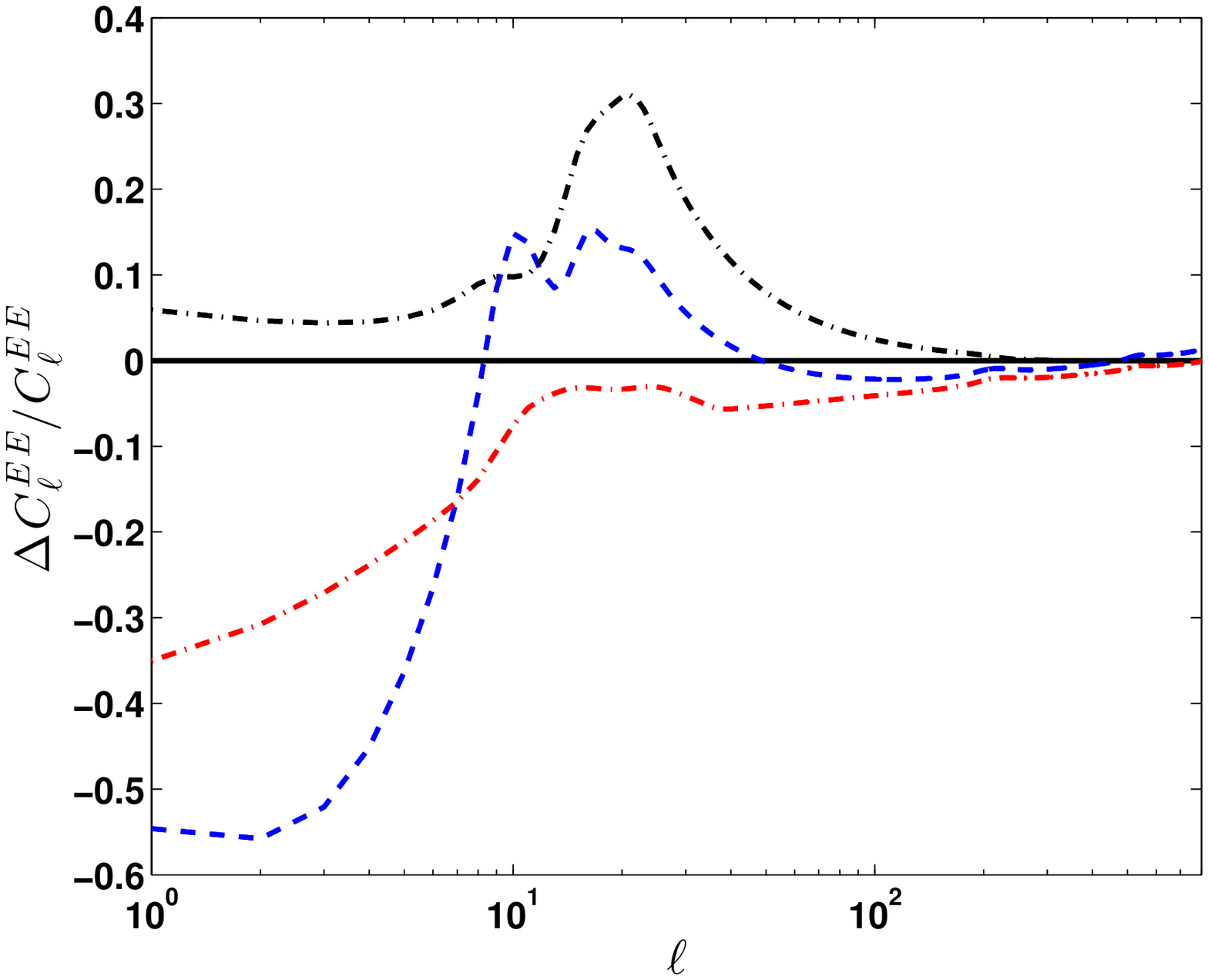}
\caption{ The EE power spectrum with 
${\cal D}^{EE}_\ell = \ell (\ell+1) C^{EE}_\ell/2\pi$. The curves' descriptions and the parameters 
are the same as in Fig. \ref{TT}. 
   }
\label{EE} 
\end{figure}

We also presents the predictions of our model for the TT, EE and BB correlations using the CAMB software. In order  to run CAMB we make an analytic template which mimics the tensor and scalar power spectra. We then play with the scale at which the transition occurs, $k_*$, and also slightly change the initial amplitudes to obtain reasonable plots. In the plots we have set  $k_*=0.0005 {\mathrm{Mpc}^{-1}}$. 

In Fig. \ref{TT} we present the temperature  power spectrum TT for both cases  $\cs_1 > \cs_2$ and $\cs_1 < \cs_2$.  As can be seen schematically,  on low multipoles the temperature  power spectrum can be reduced to match  the power deficiency  as observed  by PLANCK while on $\ell \ge 100$ the temperature power spectrum matches the well-measured results of $\Lambda CDM$.  Of course, to see whether our model provides  a better fit one has to perform a careful numerical analysis using the BICEP2 and PLANCK data. 

\begin{figure}
\includegraphics[width =  3.0in ]{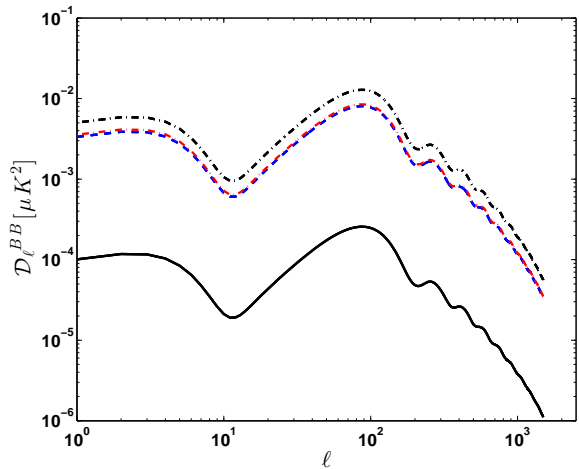}
\hspace{1cm}
\caption{The BB power spectrum with 
${\cal D}^{BB}_\ell = \ell (\ell+1) C^{BB}_\ell/2\pi$ .
The dashed and dotted-dashed curves are the same as in  previous plots. The solid black curve is for $\Lambda$CDM with $r=0.004$.
   }
\label{BB} 
\vspace{0.5cm}
\end{figure}

\begin{figure}
\includegraphics[width =  3.0in ]{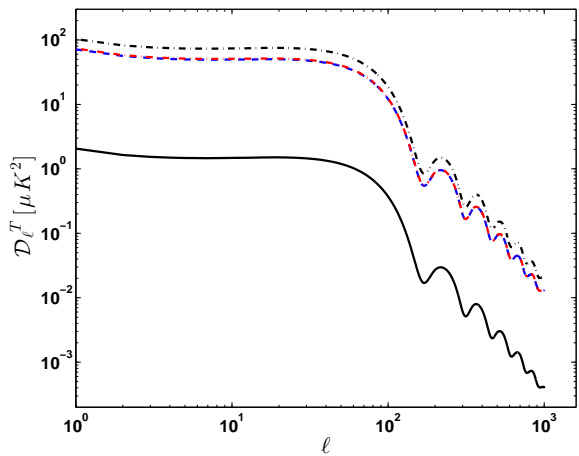}
\hspace{1cm}
\caption{Tensor power spectrum. ${\cal D}^{T}_\ell = \ell (\ell+1) C^{T}_\ell/2\pi$ .
All plots are similar to Fig.  \ref{BB}.
   }
\label{tensor-power} 
\end{figure}

In Fig. \ref{TT}  we present the EE power spectra of our model compared to the results from 
$\Lambda CDM$ with $r=0$  and $r=0.2$.  In  Fig. \ref{BB} we present the BB power spectrum of 
our model compared with the results from $\Lambda CDM$.  Finally, in Fig. \ref{tensor-power}
we present the tensor power spectrum. The general conclusion is that, compared to the $\Lambda CDM$,   one can enhance the power spectrum of tensor perturbations in our model while 
reducing the amplitude of the scalar perturbations on low multipoles. As a result, one can alleviate the tension between the PLANCK and the BICEP2 observations. Our numerical analysis are schematic, only at the level of demonstration of the validity of the idea employed. One has to perform a careful data analysis using the actual BICEP2 and PLANCK data to see whether this theory is a better fit compared to
standard $\Lambda CDM$ model.\\

To summarize, we have shown that the  tension between  the PLANCK and the BICEP2 observations can be alleviated in models of single fluid inflation in which 
the microscopical properties of the fluid such as the equation of state or the sound speed 
undergo a  jump. As we have seen, the tensor perturbations do not feel the changes in $w$ and $\cs$ directly while the changes in $w$ and $\cs$ directly affect the evolution of the scalar perturbations. Technically this originated from the non-trivial matching condition for $\calR'$
as given in Eq. (\ref{matchR2}) and the fact that, unlike the tensor perturbations, the evolution of scalar perturbations is controlled by the sound horizon incorporating the additional parameter $\cs$.
The qualitative jump in $\calP_\calR$ for the case $\cs_2 > \cs_1$ is different than the case
$\cs_2< \cs_1$ as can be seen in Figures \ref{T-radiation1} and \ref{T-radiation2}. This is attributed to the fact that in scenarios with $\cs_2 > \cs_1$ the structure of the comoving sound horizon is different the usual inflationary models in which $\calH/\cs$ is a monotonically increasing function.

In this paper we followed a phenomenological approach and did not present a dynamical mechanism 
in generating the rapid jump  in $w$ and $\cs$. As we argued before, this may be engineered if one
couples the inflaton fluid/field to other fluids/fields. The idea of a sharp waterfall during the early stage of inflation, as used in \cite{Abolhasani:2012px}, is an interesting example which can be used to engineer the jump in $\epsilon$ dynamically. Alternatively, one may look into models of brane inflation from string theory in which there are brane annihilations during inflation such as in 
\cite{Battefeld:2010rf, Firouzjahi:2010ga}.
Also one may find a natural mechanism for a sudden change
either in slow-roll parameters or the sound speed $c_s$ via fields annihilation and particle creations
\cite{Battefeld:2010vr, Barnaby:2009dd, Barnaby:2010ke, Battefeld:2013bfl}.

\section*{Acknowledgment}

We would like to W. Hu for useful discussions.  We also thank. H. Moshafi for the assistance in using CAMB.



\section*{References}

\begin{thebibliography}{}

\bibitem{Ade:2014xna} 
  P.~A.~R.~Ade {\it et al.}  [BICEP2 Collaboration],
  ``BICEP2 I: Detection Of B-mode Polarization at Degree Angular Scales,''
  arXiv:1403.3985 [astro-ph.CO].


\bibitem{Ade:2013uln} 
  P.~A.~R.~Ade {\it et al.}  [Planck Collaboration],
  ``Planck 2013 results. XXII. Constraints on inflation,''
  arXiv:1303.5082 [astro-ph.CO].

\bibitem{Contaldi:2014zua} 
  C.~R.~Contaldi, M.~Peloso and L.~Sorbo,
  ``Suppressing the impact of a high tensor-to-scalar ratio on the temperature anisotropies,''
  arXiv:1403.4596 [astro-ph.CO].

\bibitem{Miranda:2014wga} 
  V.~'c.~Miranda, W.~Hu and P.~Adshead,
  ``Steps to Reconcile Inflationary Tensor and Scalar Spectra,''
  arXiv:1403.5231 [astro-ph.CO].
  
\bibitem{Hazra:2014jka} 
  D.~K.~Hazra, A.~Shafieloo and G.~F.~Smoot,
  ``Whipped inflation,''
  arXiv:1404.0360 [astro-ph.CO].

\bibitem{Abazajian:2014tqa} 
  K.~N.~Abazajian, G.~Aslanyan, R.~Easther and L.~C.~Price,
  ``The Knotted Sky II: Does BICEP2 require a nontrivial primordial power spectrum?,''
  arXiv:1403.5922 [astro-ph.CO].

\bibitem{Smith:2014kka} 
  K.~M.~Smith, C.~Dvorkin, L.~Boyle, N.~Turok, M.~Halpern, G.~Hinshaw and B.~Gold,
  ``On quantifying and resolving the BICEP2/Planck tension over gravitational waves,''
  arXiv:1404.0373 [astro-ph.CO].


\bibitem{Miranda:2013wxa} 
  V.~Miranda and W.~Hu,
  ``Inflationary Steps in the Planck Data,''
  arXiv:1312.0946 [astro-ph.CO].
  
\bibitem{Miranda:2012rm} 
  V.~Miranda, W.~Hu and P.~Adshead,
  ``Warp Features in DBI Inflation,''
  Phys.\ Rev.\ D {\bf 86}, 063529 (2012)
  [arXiv:1207.2186 [astro-ph.CO]].

\bibitem{Contaldi:2003zv} 
  C.~R.~Contaldi, M.~Peloso, L.~Kofman and A.~D.~Linde,
  ``Suppressing the lower multipoles in the CMB anisotropies,''
  JCAP {\bf 0307}, 002 (2003)
  [astro-ph/0303636].

\bibitem{Park:2012rh} 
  M.~Park and L.~Sorbo,
  ``Sudden variations in the speed of sound during inflation: features in the power spectrum and bispectrum,''
  Phys.\ Rev.\ D {\bf 85}, 083520 (2012)
  [arXiv:1201.2903 [astro-ph.CO]].
  
  
\bibitem{Starobinsky:1992ts}
  A.~A.~Starobinsky,
  ``Spectrum of adiabatic perturbations in the universe when there are singularities in the inflation potential,''
  JETP Lett.\  {\bf 55}, 489 (1992)
  [Pisma Zh.\ Eksp.\ Teor.\ Fiz.\  {\bf 55}, 477 (1992)].

\bibitem{Leach:2001zf}
  S.~M.~Leach, M.~Sasaki, D.~Wands and A.~R.~Liddle,
  ``Enhancement of superhorizon scale inflationary curvature perturbations,''
  Phys.\ Rev.\ D {\bf 64}, 023512 (2001)
  [astro-ph/0101406].

\bibitem{Adams:2001vc}
  J.~A.~Adams, B.~Cresswell, R.~Easther,
  ``Inflationary perturbations from a potential with a step,''
  Phys.\ Rev.\  {\bf D64}, 123514 (2001).
  [astro-ph/0102236].


\bibitem{Gong:2005jr}
  J.~-O.~Gong,
  ``Breaking scale invariance from a singular inflaton potential,''
  JCAP {\bf 0507}, 015 (2005)
  [astro-ph/0504383].



\bibitem{Chen:2006xjb}
  X.~Chen, R.~Easther and E.~A.~Lim,
  ``Large Non-Gaussianities in Single Field Inflation,''
  JCAP {\bf 0706}, 023 (2007)
  [astro-ph/0611645].

\bibitem{Chen:2008wn}
  X.~Chen, R.~Easther and E.~A.~Lim,
  ``Generation and Characterization of Large Non-Gaussianities in Single Field Inflation,''
  JCAP {\bf 0804}, 010 (2008)
  [arXiv:0801.3295 [astro-ph]].



\bibitem{Joy:2007na}
  M.~Joy, V.~Sahni, A.~A.~Starobinsky,
  ``A New Universal Local Feature in the Inflationary Perturbation Spectrum,''
  Phys.\ Rev.\  {\bf D77}, 023514 (2008).
  [arXiv:0711.1585 [astro-ph]].


\bibitem{Hotchkiss:2009pj}
  S.~Hotchkiss and S.~Sarkar,
  ``Non-Gaussianity from violation of slow-roll in multiple inflation,''
  JCAP {\bf 1005}, 024 (2010)
  [arXiv:0910.3373 [astro-ph.CO]].



\bibitem{Abolhasani:2010kn}
  A.~A.~Abolhasani, H.~Firouzjahi and M.~H.~Namjoo,
  ``Curvature Perturbations and non-Gaussianities from Waterfall Phase Transition during Inflation,''
  Class.\ Quant.\ Grav.\  {\bf 28}, 075009 (2011)
  [arXiv:1010.6292 [astro-ph.CO]].



\bibitem{Arroja:2011yu}
  F.~Arroja, A.~E.~Romano and M.~Sasaki,
  ``Large and strong scale dependent bispectrum in single field inflation from a sharp feature in the mass,''
  Phys.\ Rev.\ D {\bf 84}, 123503 (2011)
  [arXiv:1106.5384 [astro-ph.CO]].

\bibitem{Adshead:2011jq}
  P.~Adshead, C.~Dvorkin, W.~Hu and E.~A.~Lim,
  ``Non-Gaussianity from Step Features in the Inflationary Potential,''
  Phys.\ Rev.\ D {\bf 85}, 023531 (2012)
  [arXiv:1110.3050 [astro-ph.CO]].


\bibitem{Achucarro:2012fd}
  A.~Achucarro, J.~-O.~Gong, G.~A.~Palma and S.~P.~Patil,
  ``Correlating features in the primordial spectra,''
  arXiv:1211.5619 [astro-ph.CO].

\bibitem{Cremonini:2010ua}
  S.~Cremonini, Z.~Lalak and K.~Turzynski,
  ``Strongly Coupled Perturbations in Two-Field Inflationary Models,''
  JCAP {\bf 1103}, 016 (2011)
  [arXiv:1010.3021 [hep-th]].

\bibitem{Avgoustidis:2012yc}
  A.~Avgoustidis, S.~Cremonini, A.~-C.~Davis, R.~H.~Ribeiro, K.~Turzynski and S.~Watson,
  ``Decoupling Survives Inflation: A Critical Look at Effective Field Theory Violations During Inflation,''
  JCAP {\bf 1206}, 025 (2012)
  [arXiv:1203.0016 [hep-th]].


\bibitem{Romano:2008rr}
  A.~E.~Romano and M.~Sasaki,
  ``Effects of particle production during inflation,''
  Phys.\ Rev.\ D {\bf 78}, 103522 (2008)
  [arXiv:0809.5142 [gr-qc]].

\bibitem{Ashoorioon:2006wc} 
  A.~Ashoorioon and A.~Krause,
  ``Power Spectrum and Signatures for Cascade Inflation,''
  hep-th/0607001; \\
  A.~Ashoorioon, A.~Krause and K.~Turzynski,
  ``Energy Transfer in Multi Field Inflation and Cosmological Perturbations,''
  JCAP {\bf 0902}, 014 (2009)
  [arXiv:0810.4660 [hep-th]].

  
  
\bibitem{Battefeld:2010rf} 
  D.~Battefeld, T.~Battefeld, H.~Firouzjahi and N.~Khosravi,
  ``Brane Annihilations during Inflation,''
  JCAP {\bf 1007}, 009 (2010)
  [arXiv:1004.1417 [hep-th]].


\bibitem{Firouzjahi:2010ga} 
  H.~Firouzjahi and S.~Khoeini-Moghaddam,
  ``Fields Annihilation and Particles Creation in DBI inflation,''
  JCAP {\bf 1102}, 012 (2011)
  [arXiv:1011.4500 [hep-th]].

\bibitem{Bean:2008na} 
  R.~Bean, X.~Chen, G.~Hailu, S.~-H.~H.~Tye and J.~Xu,
  ``Duality Cascade in Brane Inflation,''
  JCAP {\bf 0803}, 026 (2008)
  [arXiv:0802.0491 [hep-th]].
  
  
\bibitem{Battefeld:2010vr}
  D.~Battefeld, T.~Battefeld, J.~T.~Giblin, Jr. and E.~K.~Pease,
  ``Observable Signatures of Inflaton Decays,''
  JCAP {\bf 1102}, 024 (2011)
  [arXiv:1012.1372 [astro-ph.CO]].
  
\bibitem{Emery:2012sm} 
  J.~Emery, G.~Tasinato and D.~Wands,
  ``Local non-Gaussianity from rapidly varying sound speeds,''
  JCAP {\bf 1208}, 005 (2012)
  [arXiv:1203.6625 [hep-th]].

\bibitem{Bartolo:2013exa} 
  N.~Bartolo, D.~Cannone and S.~Matarrese,
  ``The Effective Field Theory of Inflation Models with Sharp Features,''
  JCAP {\bf 1310}, 038 (2013)
  [arXiv:1307.3483 [astro-ph.CO]].
  
\bibitem{Nakashima:2010sa} 
  M.~Nakashima, R.~Saito, Y.~-i.~Takamizu and J.~'i.~Yokoyama,
  ``The effect of varying sound velocity on primordial curvature perturbations,''
  Prog.\ Theor.\ Phys.\  {\bf 125}, 1035 (2011)
  [arXiv:1009.4394 [astro-ph.CO]].
  
\bibitem{Liu:2011cw} 
  J.~Liu and Y.~-S.~Piao,
  ``A Multiple Step-like Spectrum of Primordial Perturbation,''
  Phys.\ Lett.\ B {\bf 705}, 1 (2011)
  [arXiv:1106.5608 [hep-th]].


  

  
  
  
  
\bibitem{Chen:2013kta} 
  X.~Chen, H.~Firouzjahi, M.~H.~Namjoo and M.~Sasaki,
  ``Fluid Inflation,''
  JCAP {\bf 1309}, 012 (2013)
  [arXiv:1306.2901 [hep-th]].

\bibitem{Deruelle:1995kd} 
  N.~Deruelle and V.~F.~Mukhanov,
  ``On matching conditions for cosmological perturbations,''
  Phys.\ Rev.\ D {\bf 52}, 5549 (1995)
  [gr-qc/9503050].




\bibitem{Namjoo:2012xs} 
  M.~H.~Namjoo, H.~Firouzjahi and M.~Sasaki,
  ``Multiple Inflationary Stages with Varying Equation of State,''
  JCAP {\bf 1212}, 018 (2012)
  [arXiv:1207.3638 [hep-th]].


\bibitem{Abolhasani:2012px} 
  A.~A.~Abolhasani, H.~Firouzjahi, S.~Khosravi and M.~Sasaki,
  ``Local Features with Large Spiky non-Gaussianities during Inflation,''
  JCAP {\bf 1211}, 012 (2012)
  [arXiv:1204.3722 [astro-ph.CO]].



\bibitem{Barnaby:2009dd}
  N.~Barnaby and Z.~Huang,
  ``Particle Production During Inflation: Observational Constraints and Signatures,''
  Phys.\ Rev.\ D {\bf 80}, 126018 (2009)
  [arXiv:0909.0751 [astro-ph.CO]].

\bibitem{Barnaby:2010ke}
  N.~Barnaby,
  ``On Features and Nongaussianity from Inflationary Particle Production,''
  Phys.\ Rev.\ D {\bf 82}, 106009 (2010)
  [arXiv:1006.4615 [astro-ph.CO]];
  N.~Barnaby,
  ``Nongaussianity from Particle Production During Inflation,''
  Adv.\ Astron.\  {\bf 2010}, 156180 (2010)
  [arXiv:1010.5507 [astro-ph.CO]].

\bibitem{Battefeld:2013bfl} 
  D.~Battefeld, T.~Battefeld and D.~Fiene,
  ``Particle Production during Inflation in Light of PLANCK,''
  arXiv:1309.4082 [astro-ph.CO].




 



















\end{thebibliography}
\end{document}